\documentclass{eptcs}
\usepackage[pdftex]{color,graphicx}
\usepackage{times}
\usepackage{subfig}
\usepackage{longtable}
\usepackage{comment,xspace}
\usepackage{wrapfig}
\usepackage{amssymb}
\usepackage{amsmath}
\usepackage{stmaryrd}
\usepackage{mathpartir}
\usepackage{bussproofs}

\providecommand{\urlalt}[2]{\href{#1}{#2}}
\providecommand{\doi}[1]{doi:\urlalt{http://dx.doi.org/#1}{#1}}
\begin{document}

\title{Towards a Framework for Behavioral Specifications of OSGi Components } 
\author {Jan Olaf Blech
\institute{fortiss GmbH, Munich, Germany}
}

\def\titlerunning{Behavioral Specifications for OSGi}
\def\authorrunning{J.~O. Blech}

\maketitle
\begin{abstract}
We present work on behavioral specifications of OSGi components. Our behavioral specifications are based on finite automata like formalisms. Behavioral specifications can be used to  find appropriate components to interact with, detect incompatibilities between communication protocols of components and potential problems resulting from the interplay of non-deterministic component specifications. These operations can be carried out during development and at runtime of a system. Furthermore, we describe work carried out using the Eclipse based implementation of our framework.
\end{abstract}

\section{Introduction}

Traditional software component systems come -- if at all -- with a basic typing that indicates possible values at the component interface. In our work, we are extending this view to specify possible behavior of components in addition to the basic typing. We use a typing that encapsulates protocols specified by finite automata based descriptions. We present a first version of an implementation for the OSGi \cite{osgi} framework. OSGi allows dynamic reconfiguration of Java based software systems. We demonstrate a tool based approach that allows the specification of method call based communication protocols, and the formalization of creation and deletion of components during a system's lifetime. 
We check possible behavior of interacting components for behavioral compatibility including deadlocks. We can resolve possible incompatibilities by choosing options from non-deterministic behavioral specifications and -- after discovery of a potential incompatibility -- reacting inside the components accordingly.

This work describes efforts towards an operationalization and an implementation of a behavioral types framework for OSGi. It realizes parts of our vision described in \cite{isolavision12}. Unlike our work presented in \cite{blechschaetz12}, it is realized entirely using Java technology and is aimed towards the OSGi component system. A more comprehensive version of our OSGi semantics is described in a report \cite{reportosgisem}. In this paper, we primarily address protocol based behavior of components.
The new contributions of this work comprise: 
\begin{itemize}
\item A formal definition of the OSGi semantics that is suitable for the abstract view that our behavioral types provide on OSGi.
\item A first implementation of a finite automata based behavioral type system for OSGi that integrates different tools and workflows into a framework.
\item Early versions of editors and related code for supporting adaption and checking.
\item An exemplarily integration of behavioral type checkers comprising minimization, normalization and comparison. One checker has been implemented in plain Java. Additionally we have integrated a checker and synthesis tool presented in \cite{cheng2011synthesis} for deciding compatibility, deadlock freedom and detecting conflicts in non-deterministic specifications at runtime and development time. 
\item Usage scenarios (interaction protocols) of our behavioral types for OSGi at runtime and development time.
\item The modeling of an example system: a booking system to show different usage scenarios.
\end{itemize}

\subsection{Our Setting}

\begin{figure}
\centering
\includegraphics[width=0.75\textwidth,angle=0]{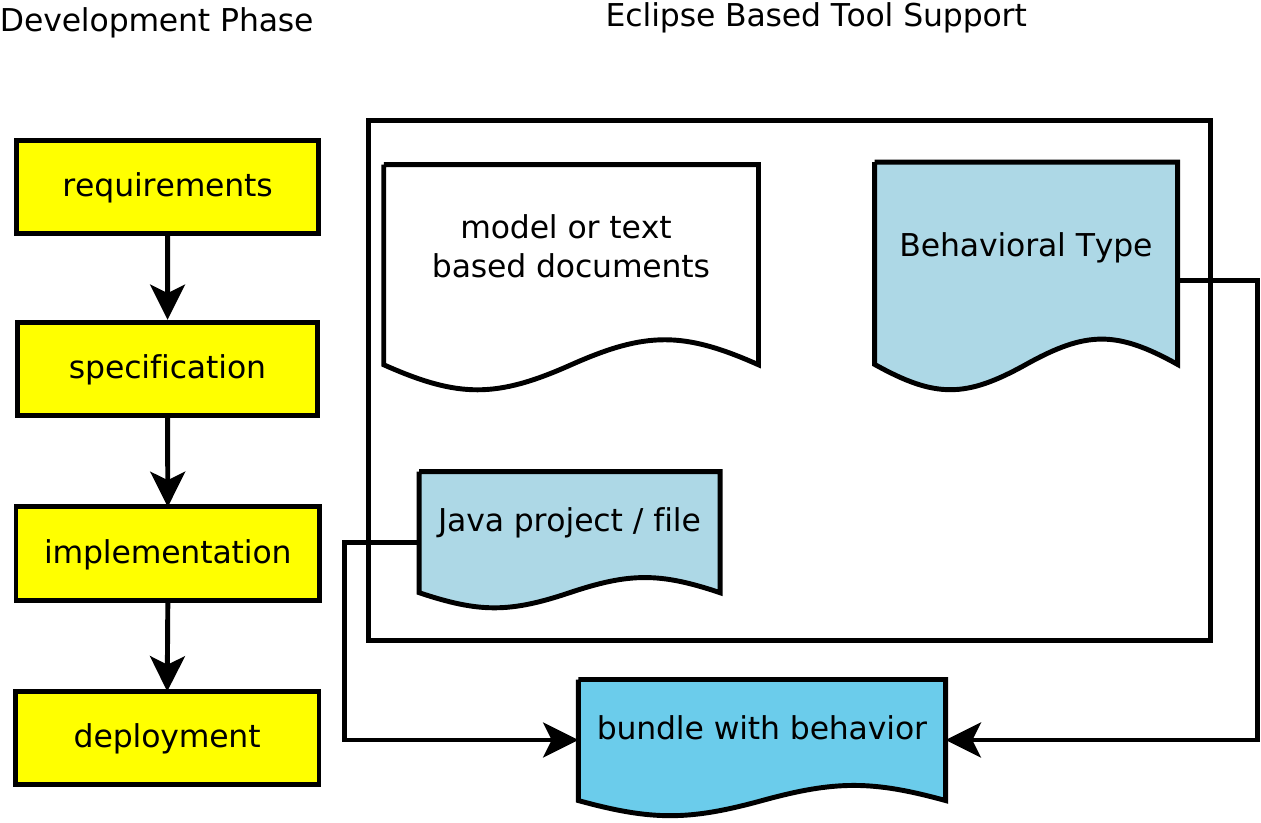}
\caption{Behavioral types at development time}
\label{fig:devchain}
\end{figure}

\begin{figure}
\centering
\includegraphics[width=0.75\textwidth,angle=0]{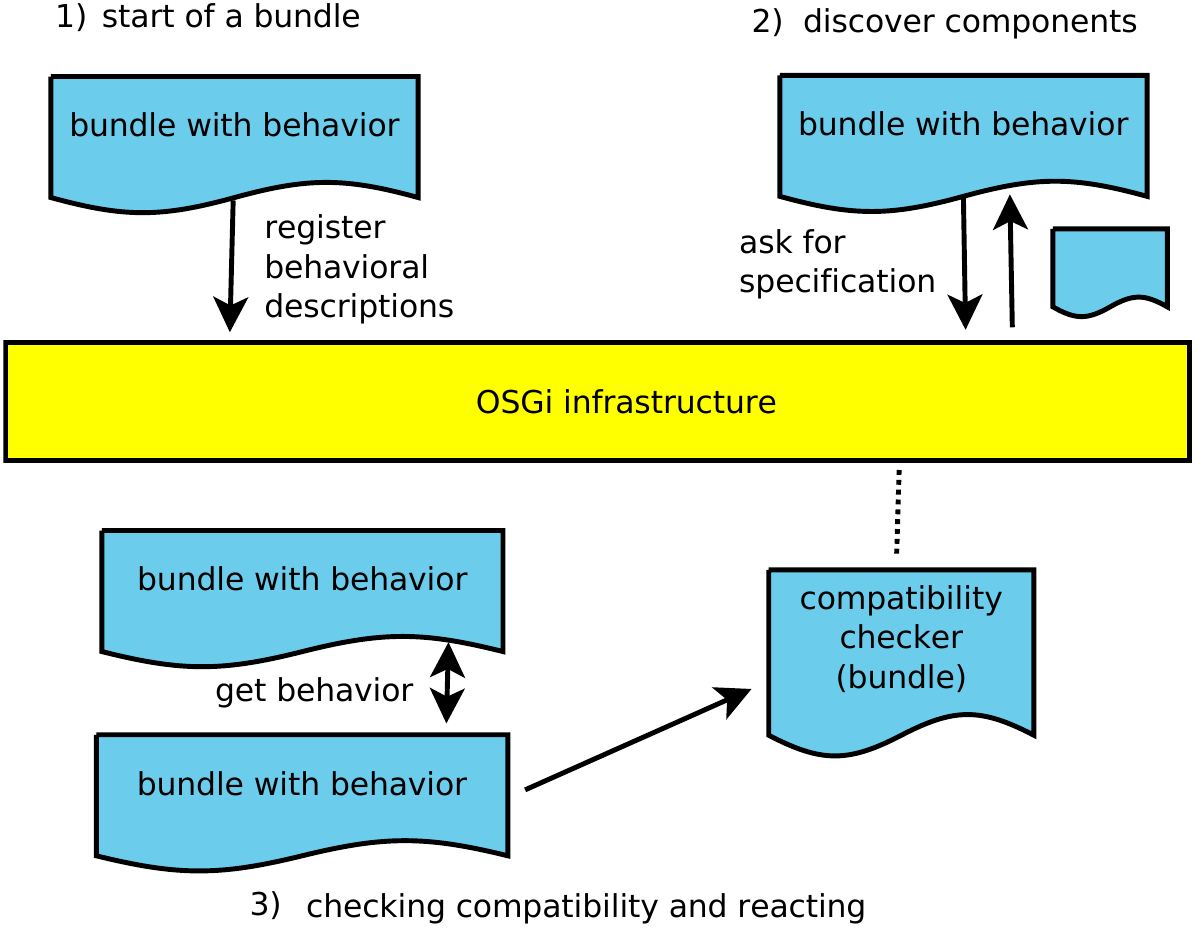}
\caption{Behavioral types at runtime}
\label{fig:behtrt}
\end{figure}

Figure~\ref{fig:devchain} shows the development chain supported by our framework.
The development of our Eclipse based development process for OSGi bundles can roughly be divided into four phases. Bundles are top-level OSGi components that aggregate classes, Java packages and deployment information. The four phases are supported by our behavioral descriptions in the following way.
\begin{itemize}
\item In the requirements and specification phase, after the component / bundle structure has been determined, one can start using our tools for behavioral types. Requirements on components and specified protocols for component interaction can be described by using the automata like specification mechanisms provided by our behavioral types. 
\item During the implementation, one creates bundles which contain the OSGi bundle information: static dependencies, classically typed interface descriptions, objects to be created at start of the bundle and Eclipse specific plugin information, e.g., extensions to the user interface. In addition to this we add our behavioral descriptions. These are given as files and can become accessible through the OSGi registration service. The OSGi registration service keeps tracks of objects / services provided by bundles and their properties.
\item At deployment and runtime of the system one has bundles including their behavioral specifications. These are 1) registered at the OSGi infrastructure and 2) can be used to discover appropriate components. Components can further use these (as shown in Figure~\ref{fig:behtrt})  to decide 3) whether and how they want to interact, to discover potential incompatibilities and ways to resolve them. Decision may be based on algorithms and tools which are provided as separate bundles.
\end{itemize}

\subsection{Related Work}
\label{sec:rw}

Interface automata \cite{AlfaroHenzinger:2001} are one form of behavioral types. Like in this work, component descriptions are based on automata. The focus is on communication protocols between components which is one aspect that we also address in this paper. Interface automata are especially aimed at compatibility checks of different components interacting at compile time of a system.  Behavioral types have also been used in the Ptolemy framework \cite{betypesptolemy} with a focus on real-time systems.

Specification and contract languages for component based systems have been studied in the context of web services. A process algebra like language and deductive techniques are studied in \cite{webcontracts}. Another process algebra based contract language for web services is studied in \cite{webcontrmath}. Emphasize in the formalism is put on compliance, a correctness guaranty for properties like deadlock and livelock freedom. Another algebraic approach to service composition is featured in \cite{conssercomp}.

JML \cite{jml} provides assertions, pre- and postconditions for Java programs. It can be used to specify aspects of behavior for Java methods.  A similar description mechanism has been used for systems specified in synchronous dataflow languages like Lustre~\cite{ColacoPouzet:2003}. Assertion like behavioral specifications have also been studied in the context of access permissions~\cite{plural}.

Behavioral types as means for behavioral checks at runtime for component based systems have been investigated in~\cite{abt}. In this work, the focus is rather put on the definition of a suitable formal representation to express types and investigate their methodical application in the context of a model-based development process. 

A language for behavioral specification of components, in particular of  object oriented systems -- but not OSGi --, is introduced in~\cite{brochjohnsen12}. Compared to the requirement-based descriptions proposed in our paper, the specifications used in \cite{brochjohnsen12}  are still relatively close to an implementation. Recent work regarding refinement of automata based specifications is, e.g., studied in \cite{behavioralrefinement}.

To the best of our knowledge, existing work does describe OSGi and its semantics only at a very high level. Other behavioral type like frameworks do not exist for OSGi up till now. A specification of the OSGi semantics based on process algebras is featured in \cite{mekontsotchinda:inria-00619233}.
Some investigations on the relation between OSGi and some more formal component models have been done in \cite{muellerse2010}. Means for ensuring OSGi compatibility of bundles realized by using  an advanced versioning system for OSGi bundles based on their type information is studied in \cite{brada09}. Aspects on formal security models for OSGi have been studied in \cite{osgisecuritybycontr}.

\subsection{Overview}
We present OSGi and our formalization of its semantics in Section~\ref{sec:osgi}. Section~\ref{sec:beht} introduces our automata based behavioral types specification mechanism. Operations on behavioral types at development and at runtime are described in Section~\ref{sec:operations} for the OSGi framework. Implementation of the framework using Eclipse / OSGi techniques is described in Section~\ref{sec:tools}. Section~\ref{sec:booksys} exemplifies the use of behavioral types and its operations for a booking system and a conclusion is given in Section~\ref{sec:conc}.

\section{OSGi and its Semantics}
\label{sec:osgi}

\begin{figure}[t]
\centering
\includegraphics[scale=0.75]{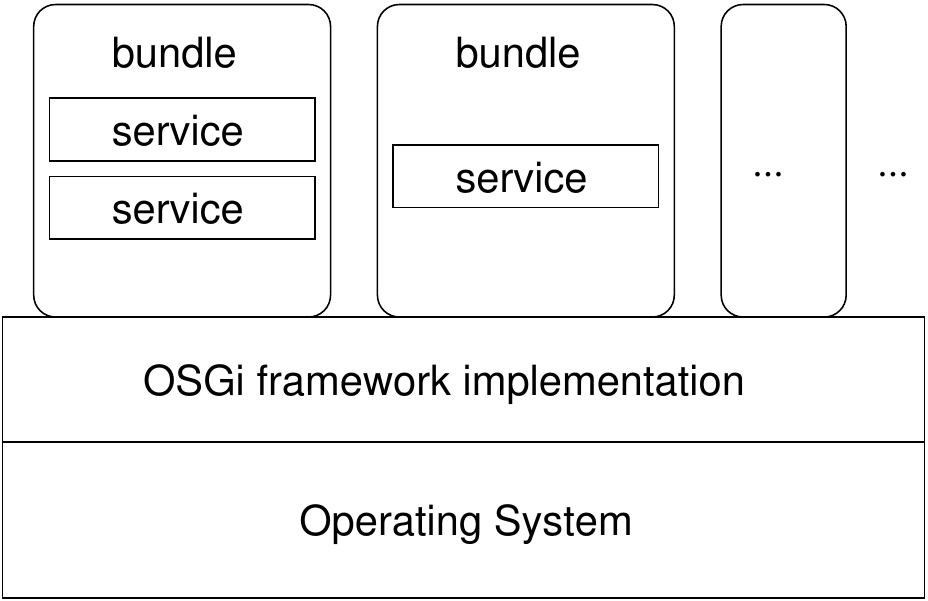}
\caption{OSGi framework}
\label{fig:osgiexample}
\end{figure}
We present an overview on OSGi following our description in \cite{isolavision12} and present a formalization of the semantics based on our more detailed report \cite{reportosgisem}.

The OSGi framework is a component and service platform for Java. It allows the aggregation of Java packages and classes into bundles (cf. Figure~\ref{fig:osgiexample}) and comes with additional deployment information. The deployment information  triggers the registration of services for the OSGi framework.
Bundles provide means for dynamically configuring services, their dependencies and usages. OSGi bundles are used as the basis for Eclipse plugins but also for embedded applications including solutions for the automotive domain, home automation and industrial automation. Bundles can be installed and uninstalled during the runtime. For example, they can be replaced by newer versions. Hence, possible interactions between bundles can in general not be determined statically.

Bundles are deployed as .jar files containing extra OSGi information. This extra information is stored in a special file inside the .jar file. Bundles generally contain a class implementing an OSGi interface that contains code for managing the bundle, e.g., code that is executed upon activation and stopping of the bundle. 
Upon activation, a bundle can register its services to the OSGi framework and make it available for use by other bundles. Services are implemented in Java. The bundle may itself start to use existing services. Services can be found using dictionary-like mechanisms provided by the OSGi framework. Typically one can search for a service which is provided using an object with a specified Java interface. 

In the context of this paper, we use the term OSGi component as a subordinate concept for bundles, objects and services provided by bundles.

The OSGi standard only specifies the framework including the syntactical format specifying what bundles should contain. Different implementations exist for different application domains like Equinox\footnote{\url{http://www.eclipse.org/equinox/}} for Eclipse, Apache Felix\footnote{\url{http://felix.apache.org/site/index.html}} or Knopflerfish\footnote{\url{http://www.knopflerfish.org/}}. If bundles do not depend on implementation specific features, OSGi bundles can  run on different implementations of the OSGi framework.

\subsection*{A Method-Call Semantics}
\label{sec:methodcallsem}
In the following we provide a formal semantics for OSGi. We concentrate on capturing behavior originating from method calls between different bundles and objects. Memory and exchange of data between these bundles and objects is not taken into account. Thus, we provide an overapproximation -- in the sense of possible behavior -- and abstraction of a real system. 

\paragraph{Object and method definitions}
\label{def:semmeth} 
An object is defined as a tuple $(m_0,...,m_n)$ comprising constructor and method definitions $m_0,...,m_n$. Since we incorporate constructors into this tuple it cannot be empty.

The semantics of an object is given by  the semantic interpretation of its methods and its object state. 
The semantics of a method is giving by an automaton $(L,E,l_0)$ comprising a set of locations $L$ an initial location $l_0 \in L $ and edges $E= (l_i,M,l_j)$ between locations. 
An addition to source and target location  $l_i$ and $l_j$ an edge comprises a set (can be ordered) of method calls and special calls $M$. These can be tuples $(m,o,b) \in M$ comprising a method definition $m$ of an object $o$ that is associated with a bundle $b$. Furthermore, $M$ can contain special calls for:
%\begin{itemize}
%\item 
adding and removing bundles, and
%\item 
creating and deleting objects.
%y\end{itemize}
Each transition from $E$ represents an action that is atomic or non-terminating to the method but not to the OSGi system. It can represent a memory update, but also other method calls. A method call can itself trigger a non-terminating method in the same or in other objects. Therefore a transition does not necessarily represent a terminating operation.

An object state is a set of tuples 
%\begin{center}
$\{ (m_n,l_{n_i},id_{n},cs_n), ... , (m_p,l_{p_j},id_{p},cs_p) \}$ 
%\end{center}
comprising active method status states $(m_n,l_{n_i},id_n,cs_n), ... , (m_p,l_{p_j},id_p,cs_p)$. Each tuple represents a method call, consisting of a method definition, its actual locations, an identifier $id$ and a call state $cs$.

The call state is part of an active method status state. It is a set of method definitions and method id plus status information for which the active method is waiting to return. The id is used to distinguish different calls to the same method.

\paragraph{Bundles}
\label{def:bundle}
From an operational semantics point of view, bundles aggregate objects into units that are enumerated in the OSGi system and can be loaded and removed during runtime by user commands or from other bundles.
A bundle is a set of objects $\{ o_{activator}, ... \}$ comprising an object $o_{activator} = (...,m_{start},m_{stop},...)$ which is created on activation. It comprises two distinguished methods $m_{start},m_{stop}$ which are called during activation and deactivation.

In an implemented OSGi system, the $o_{activator}$  object has to implement the \\ {\tt BundleActivator} interface defined in {\tt org.osgi.framework}. It comprises two methods with signatures:
\begin{center}
{\tt void start(BundleContext context) throws java.lang.Exception}
\end{center}
and
\begin{center}
{\tt void stop(BundleContext context) throws java.lang.Exception}
\end{center}

The semantical definition of bundle states aggregates its object states.
A bundle state is defined as a set of object states $\{ s_{o_i},...,s_{o_k} \}$ %
for object states $s_{o_i},...,s_{o_k}$.
Like for bundles, objects, and methods, we distinguish between a system state and a system definition -- capturing a systems architecture. Both can change during the lifetime of a system. A standard OSGi system has one (as, e.g., in the Equinox framework implementation) or more bundles which are active at startup.

\paragraph{OSGi systems and OSGi system states}
An OSGi system is a set of bundles. It comprises a distinguished bundle $b_{init}$ which is activated at start-up.
Analog to object and bundle state, we define an OSGi system state.
A system state is defined as a set of bundle states  $\{ s_{b_i},...,s_{b_k} \}$ for bundle states $s_{b_i},...,s_{b_k}$.
The initial state of an OSGi system comprises the start of the $start$ method in the activator object of the initial bundle.
The initial state of an OSGi system is defined as $s_{init} = \{s_{b_{init}} \}$ with $s_{b_{init}} =  \{ o_{activator} \}$ and $o_{activator} = \{ (m_{start},l_{{start}_0},0,\emptyset) \}$.

\paragraph{Dynamic architecture of OSGi systems}
\label{sec:dynoperations}

An important aspect of our formalization is the impact on OSGi operations that can change the structure of OSGi systems. Such operations can be triggered by OSGi methods themselves, 
e.g., comprising adding and removing objects and bundles. 
Another option is to perform these operations by a command line interface (e.g., starting Eclipse with the console option using Equinox) at runtime on the OSGi framework.
Here, we distinguish the following structure changing operations on OSGi systems:
%\begin{itemize}
%\item 
Starting / loading a system,
%\item 
adding a bundle and activating it, 
%\item 
removing a bundle (and deactivating it),
%\item 
adapting a bundle and its services,
%\item 
closing / removing a system.
%\end{itemize}
Characteristic for these operations is the fact that new behavior  becomes possible or is removed at runtime of the OSGi system. Thus, the semantics of an OSGi system and possible events can in general not be determined statically at the start of a system.

\paragraph{State transitions in OSGi}
\label{sec:statetrans}
State transitions can modify both, structure of a system and the state of objects, bundles and a system.
They are made up from local transitions appearing within methods and from handling terminated methods. In general state transitions are highly non-deterministic and define a relation of 
\begin{center}
previous system state $\times$ previous system definition $\times$ \\
next system state $\times$ next system definition
\end{center}

For an OSGi system $S = \{ ...,b,...\}$: We regard the system state $s = \{ ... , s_b ,...   \}$  with $s_o \in s_b $ and $ (m,l_i,id,cs) \in s_o$. From here,  the following basic state transition cases can be distinguished:
 
\begin{itemize}
\item Calling a method $m'$ of object $o'$ from a bundle $b'$:  We regard a transition  $(l_i,M,l_j) \in o$ with $ o \in b$. The following steps are performed.
\begin{enumerate}
\item
The step can be performed under the preconditions that $(m',o',b') \in M$ and $o'$ and $b'$ exist in $S$.
\item
$cs$ is updated by adding the method call indicating its bundle, object and id.
\item
A new element $(m',l'_0,id',\emptyset)$ is added to the object state where $m'$ belongs to $o'ç$. $id'$ is a new identifier for the method $m'$.
\end{enumerate}
\item Executing a method step:  We regard a transition  $(l_i,M,l_j) \in o$ with $ o \in b$.
\begin{enumerate} 
\item
The step can be performed under the precondition that $cs = \emptyset$.
\item 
$s_o$ is updated as $s'_o = s_o / (m,l_i,id,cs) \cup \{ (m,l_j,id,cshandle(M)) \}$. Thus, \\ $(m,l_i,id,cs)$ is removed and $(m,l_j,id,cshandle(M))$ is added instead. 

$cshandle$ transforms $M$ into a representation that indicates which methods have been called and keeps track of their ids. Furthermore, $cshandle$ takes care of special operations that modify the system definition.
\end{enumerate}
\item Returning from a method call:  Any method status state with $cs = \emptyset$ and no edge that may lead to a possible succeeding state can be processed in the following way:
\begin{enumerate}
\item The method status state is removed.
\item The call state of any method that $m$ has called is updated such that the entry for the $m$ call is removed.
\end{enumerate}
\end{itemize}

Furthermore, the following operations are handled:
\begin{itemize}
\item Adding a bundle : The $cs$ from any object state $s_o $ with  $ (m,l_i,id,cs) \in s_o$ can contain a special operation (denoted: {\sf add bundle $b'$}) for adding a bundle $b'$ and changing the system definition from $S$ into $S' = S \cup \{ b' \}$.
\item Removing a bundle: 
The $cs$ from any object state $s_o $ with  $ (m,l_i,id,cs) \in s_o$ can contain a special operation for removing a bundle $b'$ (denoted: {\sf remove bundle $b'$}) and changing the system definition from $S$ into $S' = S / \{ b' \}$.
\item Creating an object. The $cs$ from any object state $s_o $ with  $ (m,l_i,id,cs) \in s_o$ can contain a special operation (denoted: {\sf create object $(o',b)$}) for adding an object $o'$ and changing a bundle definition $b \in S$ to $b' = b \cup \{ o' \}$. The  system definition is, thus, changed from $S$ into $S' = S / b \cup \{ b' \}$.
\item Deleting an object:  The $cs$ from any object state $s_o $ with  $ (m,l_i,id,cs) \in s_o$ can contain a special operation (denoted: {\sf delete object $(o',b)$}) for deleting an object $o'$ and changing a bundle definition $b \in S$ to $b' = b / o' $. The  system definition is, thus, changed from $S$ into $S' = S / b \cup \{ b' \}$.
\end{itemize}

\paragraph{Key characteristics of the OSGi semantics}
The method call semantics described above features some key-characteristics of OSGi: 
\begin{itemize}
\item State transitions in bundles and objects are triggered out of the bundles and objects themselves. They only involve the component where they originate from and components that are interacted with during a state transition. The rest of the system remains untouched with respect to the underlying abstractions of our semantics.
\item Different components run asynchronously as long as there is no method call between them.
\item Method calls provide synchronization points between components.
\item Method calls are blocking. 
\end{itemize}

\section{Behavioral Types as Specification Mechanism}
\label{sec:beht}
Our framework essentially supports finite automata for specifying expected incoming, potential outgoing method calls, the creation and deletion of components during a time span and other events that may occur in the lifetime of a system. A component's behavior can be specified by one or multiple automata each one describing a behavioral aspect. Formally, we have an alphabet of labels $\Sigma$, a set of locations $L$, an initial location $l_0$ and a set of transition edges $E$ where each transition is a tuple $(l,\sigma,l')$ with $l,l' \in L$ and $\sigma \in \Sigma$. These are aggregated into a tuple to form a behavioral specification: 
\begin{center}
$(\Sigma, L,l_0,E)$
\end{center}
This view abstracts from the specifications given in Section~\ref{sec:osgi}. Our intention is to define interaction protocols or some aspects of them like the expected order of incoming and outgoing method calls for a component. Specifications for different components are independent of each other as long as there is no method call (e.g., indicated by the same label name) in the specifications.

\begin{figure}
\centering
\includegraphics[scale=0.85]{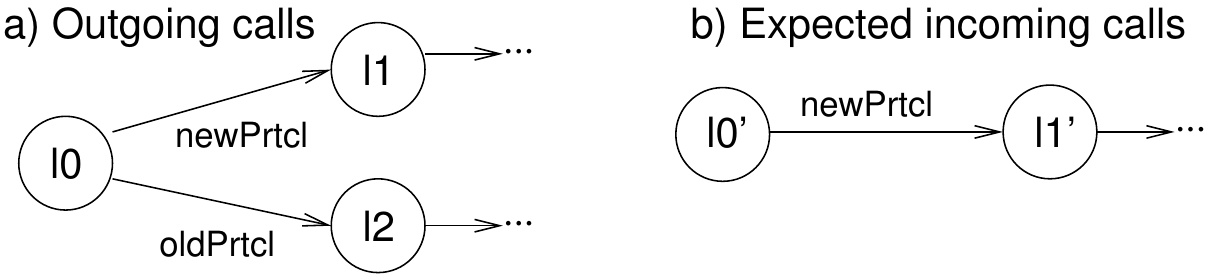}
\caption{Supporting different protocol versions}
\label{fig:oldnewprot}
\end{figure}

\paragraph{Example: Two components interacting}
Specifications can be used for different behavioral aspects. Figure~\ref{fig:oldnewprot} shows two excerpts of automata for outgoing and expected method calls from two different component specifications: 
\begin{center}
{\small $(\{newPrtcl, oldPrtcl, ... \}, \{l0,l1,l2,...\}, l0, \{(l0,newPrtcl,l1),(l0,oldPrtcl,l2) ,...\})$ \\
\vspace{0.2cm}
and \\ 
\vspace{0.2cm}
$(\{newPrtcl, ... \}, \{l0,l1,...\}, l0, \{(l0,newPrtcl,l1) ,...\})$ }
\end{center}
Here, the first component can do two different method calls in its initial state: {\sf newPrtcl, oldPrtcl}. The second component expects one method call {\sf newPrtcl} in its initial state. 
In this case both components may interact with each other, if both components use the {\sf newPrtcl}.

\section{Checking Compatibility and Making Components Compatible}
\label{sec:operations}
We describe operations that can be used at development and at runtime of a system. The operations use behavioral types from Section~\ref{sec:beht}. Furthermore, we briefly describe the handling of potential incompatibilities discovered by type comparison at runtime within a software system.

\subsection{Simple Behavioral Type Checking}
\label{sec:deccomp}
We have developed and implemented different operations for handling and comparing behavioral types, for deciding compatibility and for deadlock freedom.

Simple comparison for equality of types and comparison for refinement between two automata based specifications involves the following steps.
\begin{itemize}
\item A basis for the comparison of two types is the establishment of a set of semantical artifacts (e.g., method calls) that shall be considered. The default is to use the union of all semantical artifacts that are used in the two types. Comparison for refinement is achieved by eliminating certain semantical artifacts from this set. For consistency this also requires eliminating associated transitions from the types  or, depending on the desired semantics, replacing an edge with an empty or $\tau$ label.
\item It is convenient to complete specifications for further comparison: Specification writer may only have specified method calls or other semantical artifacts that trigger a state change. Here, we automatically add an error location. We collect possible labels and for locations that do not have an edge for a label leading to another location indicating a possible semantical artifact, we add edges with the missing label to the error location.
\item In case of specifications which have been completed and that have no locations with two outgoing edges with the same labels, we perform a minimization of automata based specifications. This way, we merge locations and get rid of unnecessary complexity automatically.
\item Normalization of automata based specifications. This, involves the ordering of edges and in some cases locations with respect to the lexicographic order of their labels / location names.  
\item Checking for equality involves the checking of equality of the labels on edges. Optionally, one can also consider the equality of location names of an automaton. Location names may imply some semantics but in our standard settings they only serve as ids. When location names serve only as ids, we  construct a mapping between location names of the two automata involved in the comparison operation.
\end{itemize}
These operations have been implemented in Java.  They do not need additional tools or non-standard plugins.

\subsection{Deciding Compatibility and Deadlock-Freedom }

In addition to the operations described in Section~\ref{sec:deccomp} we have adapted a SAT and game-based tool -- VissBIP  presented in \cite{cheng2011synthesis} -- to serve as a compatibility and deadlock checker for our behavioral types for OSGi.
Our framework uses VissBIP to support the checking of the following properties:
\begin{itemize}
\item Deadlocks checking: deadlocks resulting from potential sequences of method calls can be detected.
\item Compatibility: A component anticipating a certain behavior of incoming method calls matches potential behavior of outgoing method calls by other components.
\end{itemize}

VissBIP uses a simplified version of the BIP semantics \cite{bip1}. A system comprises concurrent automata with labeled edges. The automata synchronize with each other by performing edges with the same labels in parallel.
Otherwise, the default case is that automata do not synchronize with each other. 
For comparing method call based behavioral specifications we use VissBIP  on specifications that comprise expected incoming and outgoing method calls of components. In OSGi synchronization between components happens only when one component calls a method of the other component as indicated in the behavioral specification and the OSGi semantics. On the VissBIP side this corresponds to same labels in the automata that represent the behavior.
In addition to the label compatibility checking, VissBIP is able to perform the introduction of priorities.

\subsection{Runtime Adaption of Systems}
One way of runtime adaption is the reaction to potential deadlocks or incompatibilities.
Recall Figure~\ref{fig:oldnewprot}: it shows behavioral specifications of two components which intend to communicate with each other. Possible outgoing method calls of one component and expected incoming method calls of the other component are shown. It can be seen that the first component is able to communicate using two different protocols: one starts by calling an initialization method {\sf newPrtcl}, the other one starts by calling an initialization method {\sf oldPrtcl}. The other component expects the {\sf newPrtcl} call. 

When we give these two specifications to VissBIP, it will return a list of priorities where the {\sf newPrtcl} edge is favored over the {\sf oldPrtcl} edge in the first specification. In a Java implementation the first component can use this to dynamically decide at runtime which protocol to use. 
\begin{itemize}
\item 
First, the component loads its own behavioral specification and the specification of the expected method calls of the second component. Technically, we support loading files and the registration of models as properties / attributes of bundles as provided by the OSGi framework.
\item 
Next, we invoke VissBIP or another checking routine. Passing the behavioral specifications as parameters.
\item
The checking routine gives us a list of priorities. In the Java code we have a switch statement as a starting point for handling the different protocols. We check the priorities and go to the case for the appropriate protocol.
\end{itemize}
Thus, in addition to deadlock detection, we can use behavioral specifications for coping with different versions of components and desired interacting protocols.

\subsection{Component Discovery at Runtime}
A central feature of our behavioral descriptions for OSGi components is registering them to a central OSGi instance. In order to inform other components of the existence of a bundle with behavioral offers and needs, we register its behavioral properties using the OSGi service registry belonging to a {\tt BundleContext} which is accessible for all bundles in the OSGi system:
\begin{verbatim}
  registerService(java.lang.String[] clazzes,
     java.lang.Object service,
     java.util.Dictionary<java.lang.String,?> properties)
\end{verbatim}
Here, we register a collection of behavioral objects as properties for a service representing a bundle under a String based key. In our framework, we register a collection of behavioral models as "BEHAVIOR". The behavioral models are loaded from XML files that are integrated into the bundle. The behavioral models come with meta information which identify the parts of the behavior of a bundle which they describe. The service itself is represented as an object. Additional interface information is passed using the {\tt clazzes} argument.

\section{Tool Support during Development and at Runtime}
\label{sec:tools}
The features described in this paper have been implemented in Eclipse. Our framework offers the following ingredients and is build using the following concepts:
\begin{itemize}
\item EMF/.ecore based meta model of behavioral descriptions for easy interactions with other Eclipse based tools. 
Each specification is associated with a description which classifies what is actually specified, e.g., incoming method calls, outgoing method calls, component creation and deletion or something mixed.
\item Editors for behavioral descriptions. Figure~\ref{fig:editors} shows a screenshot of an editor for automata based specifications. 
\begin{figure}[t]
\centering
\includegraphics[width=0.7\textwidth,angle=-90]{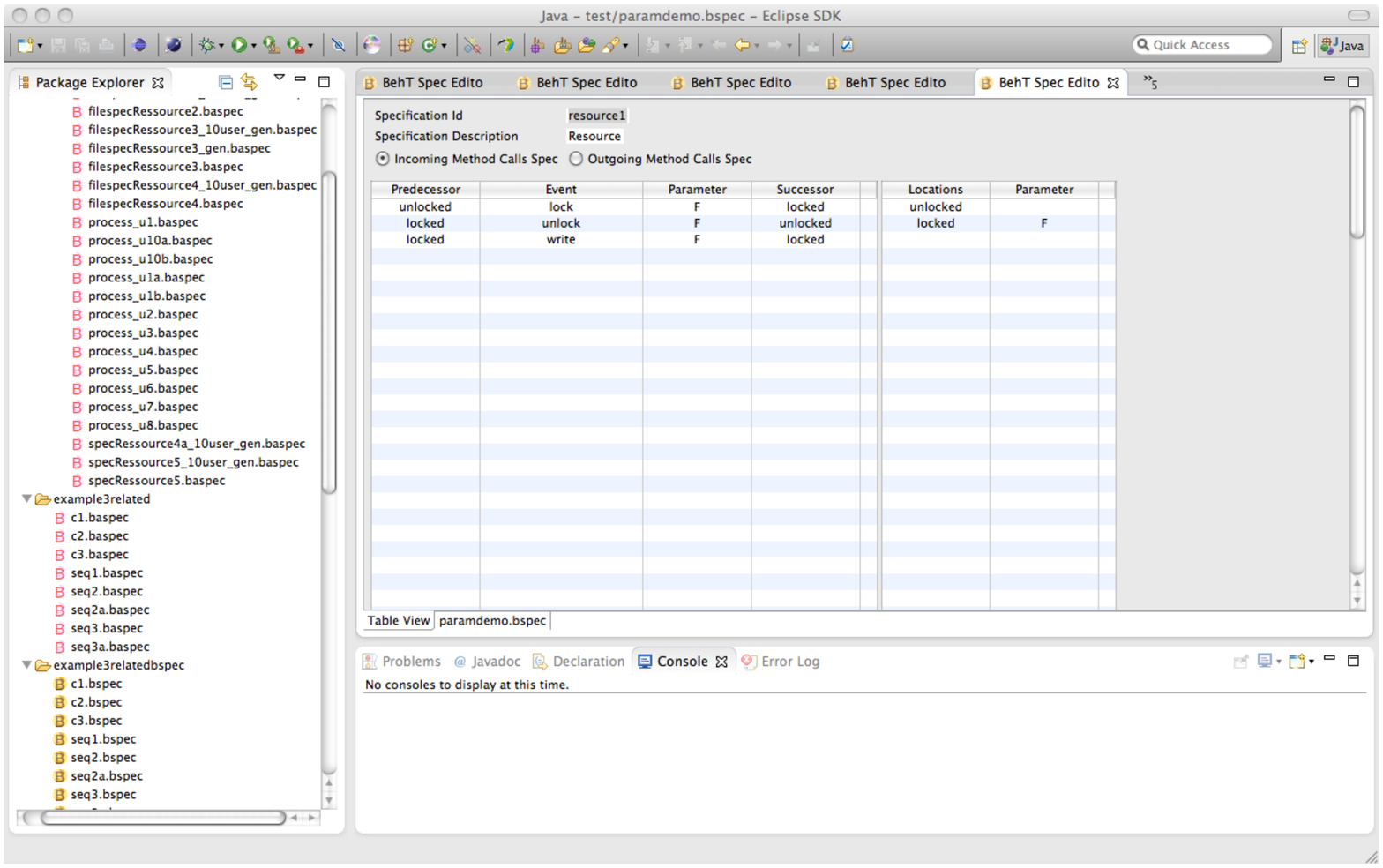}
\caption{Specifying behavior using our editors in Eclipse}
\label{fig:editors}
\end{figure}
\item Other operations like abstractions, minimization and comparison of behavioral types (some of them are described in Section~\ref{sec:deccomp}) are implemented. They can be  used by referencing one of our plugins and can be extended.
\item An integration of the VissBIP checker as Eclipse plugin / OSGi bundle and transformations for using it with our behavioral types are offered.
\end{itemize}

While the editors are only invoked at development time, the specifications and operations on them are used both: at development time and at runtime of the system for reconfiguration. At development time, they are invoked using the Eclipse front-end. At runtime, they are invoked using method calls to plugins that realize the operations and are deployed as OSGi bundles.

\section{Behavioral Types for a Booking System}
\label{sec:booksys}
We present the use of behavioral types to highlight some features and usages of our work on an example: a flight booking system.

\begin{figure}
\centering
\includegraphics[width=0.5\textwidth]{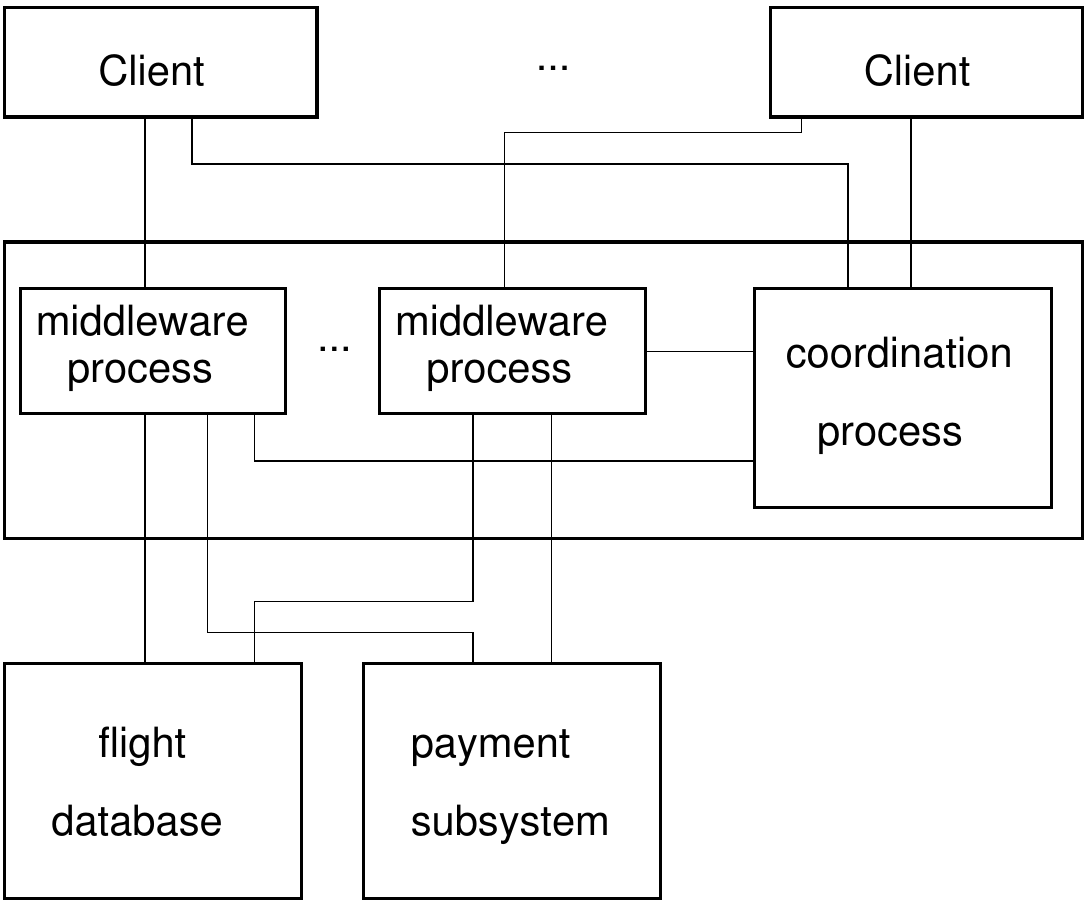}
\caption{Components of our flight booking system}
\label{fig:fbcomps}
\end{figure}

Figure~\ref{fig:fbcomps} shows the main ingredients of our flight booking system. Clients are served by middleware processes which are created and managed by a coordination process. Middleware processes use concurrently a flight database and a payment system. The described system is an example inspired by realistic systems where the middleware is implemented using Java/OSGi. In addition to the middleware components we describe databases and parts of the frontend using our behavioral types to make checks of these parts possible. 

The following means of behavioral interaction can be distinguished:
\begin{itemize}
\item {\bf Component calls between methods / communication protocol} In our flight booking system, a client can call a coordination process and middleware processes. Middleware processes can call methods providing access to the flight database and the payment subsystem. The method calls need to respect a distinct protocol which can be encoded using our behavioral types.
\item {\bf Creation and deletion of new components} The coordination process creates and removes middleware process such that there is one process per client. Providing support for analysis of such dynamic aspects is a long term goal for our behavioral types but not in the scope of this work.
\item {\bf Concurrent access to shared resources}
Middleware processes perform reservations, cancellations, rebookings, seat reservations and related operations on the flight database. These operations do require the locking of parts of the data while an operation is performed. For example, during a seat reservation a certain amount of the available seats in an aircraft is locked so that a customer can chose one without having to fear that another customer will chose the same seat at the very same time. In the current state we are able to provide some behavioral types support here.
\end{itemize}

\begin{figure}
\centering
\includegraphics[width=0.7\textwidth]{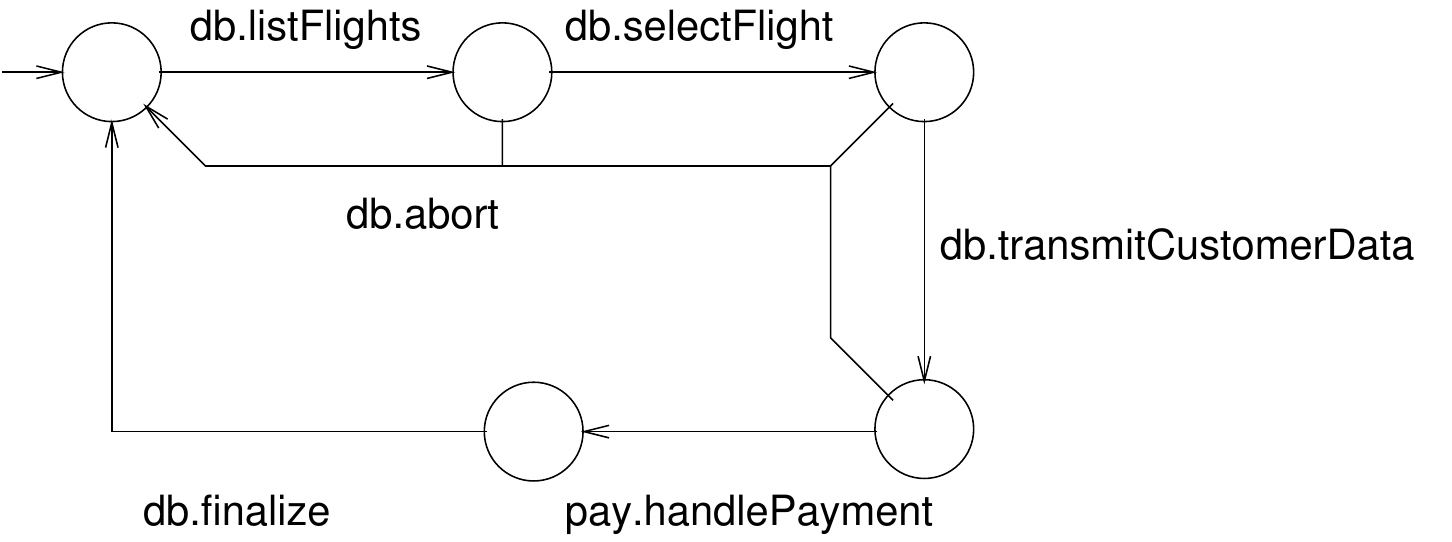}
\caption{Outgoing method calls of a middleware process}
\label{fig:commprot1}
\end{figure}

\paragraph{Example: Specification of outgoing method calls of a middleware process}
Specifications of possible expected incoming and potential outgoing method calls give information about a communication protocol that is to be preserved. Typically different interaction sequences are possible, especially since we are dealing with abstractions of behavior.
In the booking system, a middleware process communicates with a flightdatabase (db) and the payment system (pay). The expected order of method calls for a flight booking to these systems is shown in Figure~\ref{fig:commprot1}.
The figure shows only an excerpt of the possible states and transitions. In addition to this, the initial state allows the start of a seat reservation process and a cancellation process. Moreover,
Figure~\ref{fig:commprot1} shows only the state changing method calls of the behavioral specification of the booking process. Our real behavioral specification completely lists all possible method calls in each state. This way, we can further analyze compatibility issues for example with database systems that do not support all possible method calls of a middleware process.  

In comparison to the outgoing method calls of a middleware process, the incoming method call specification is much simpler: A constructor call is performed by the coordination process upon initialization. After that, the communication with the client is done using a webserver interface -- comprising method calls that send raw request data to the middleware process and return raw response data that trigger, e.g., displaying selected flights by the client -- where no states in the communication process can be distinguished. 

\paragraph{Example: Specification of database elements}
Access to our database is done using method calls to a database process and is formalized using our automata based specification formalisms. The method calls result in locking and unlocking database elements. Seat reservation in a flight requires that a certain partition of the available seats is blocked during the selection process so that a client can make a choice.

\begin{figure}
\centering
\includegraphics[width=0.75\textwidth]{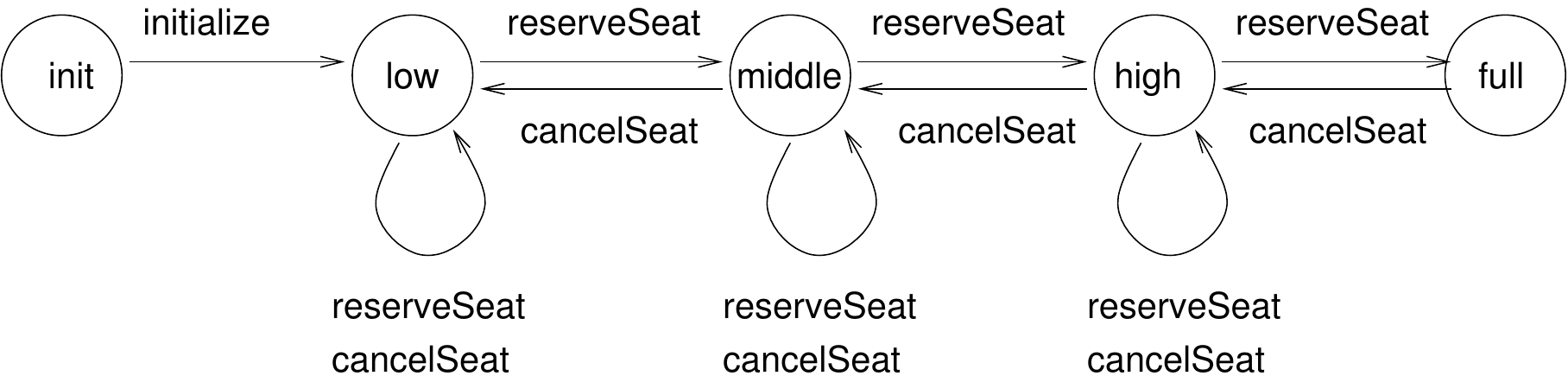}
\caption{Behavioral model for seat reservation of a flight}
\label{fig:seatres}
\end{figure}

Figure~\ref{fig:seatres} shows our behavioral model of seat reservation for a single flight. Different loads are distinguished: low means that many seats are still available, while high means that only a few seats are available. The full state indicates that no additional seat reservations can be made, only cancellations are possible. The model is an abstraction of the reality since instead of treating each seat -- potentially hundreds of available seats -- independently we only distinguish their partitioning into four equivalence classes: low, medium, high and full.

%---JO

\paragraph{Example: Database elements and deadlocks}
Access to the flight database can result in deadlocks. The model from Figure~\ref{fig:seatres} can serve as a basis for deadlock analysis. Consider the scenario shown in Figure~\ref{fig:concseatres}:
\begin{figure}
\centering
\includegraphics[width=0.7\textwidth]{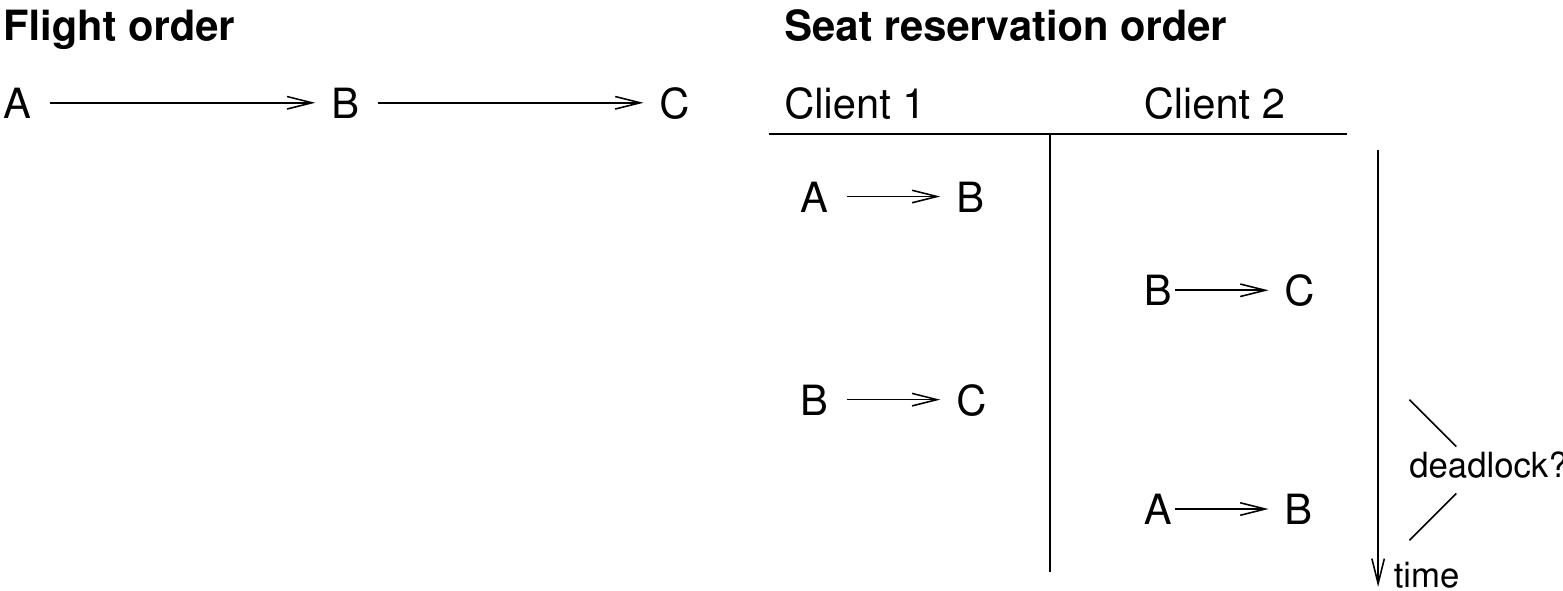}
\caption{Concurrent seat reservation on two flights}
\label{fig:concseatres}
\end{figure}
For each flight a different instance of the seat reservation model exists. Given three airports A, B and C: Suppose two people -- person 1 and person 2 -- want to fly from A to C via B. Seats for two flights need to be reserved: from A to B and from B to C. It is not desirable to reserve a seat from B to C if no seat is available for the flight to A to B. Otherwise, it might not be desirable to fly from A to B if no seat is available for the flight from B to C. 

During the seat reservation a deadlock may occur: If person 1 reserves the last seat for the A to B flight before doing reservations for the B to C flight and person 2 reserves the last seat for the B to C flight before a seat reservation for the A to B flight a deadlock may occur, which may result in the cancellation of both journeys although one person could have taken the journey.

If it is known before to the seat reservation system that person 1 and person 2 will fly from A to C -- which is a reasonable assumption given the fact that they have entered their desired start and end destination into the system --  we are able to detect such deadlocks. They can occur if both behavioral models of the seat reservation system are already in the high state -- given that no other participants are doing reservations at this time we may also take compensating actions. 

\paragraph{Evaluation}
Modeling of the flight booking system has been carried out in several versions with several degrees of detail in our behavioral types framework plugins. Behavioral models are described as independent files. We have used our implemented operations on these files. Compatibility and deadlock checking can be performed without problems for several components interacting together. For our compatibility checking, we do not use all specifications of the entire system together but pick those that are relevant for a certain communication aspect.

\section{Conclusion}
\label{sec:conc}
We presented a first version of a framework for behavioral types for OSGi systems. In this paper, the main focus is on the OSGi semantics, the specification of behavior and checking the compatibility  of specifications. Handling and reacting to  specifications at runtime is another topic. We have described our implementation and its architecture.

So far, we are concentrating on Eclipse / OSGi systems. Other application areas for the future comprise 1) work towards behavioral types for distributed software services 2) work towards real-time embedded systems. This might require leaving the Java / OSGi setting, since these applications typically involve C code which communicates directly with -- if at all -- an operating system. There is, however, work on extensions for real-time applications of OSGi using real-time Java (e.g., \cite{donsez12}). Additional specification formalisms and the integration of new checking techniques are another challenge. 

\bibliographystyle{plain}

\begin{thebibliography}{99}



\bibitem{AlfaroHenzinger:2001}
L. de~Alfaro, T.A.  Henzinger.
\newblock Interface automata.
\newblock  Symposium on Foundations of Software Engineering, ACM , 2001. 
\newblock \doi{10.1145/503271.503226}

\bibitem{donsez12} J. C. Am{\'e}rico, W. Rudametkin, and D. Donsez.
\newblock Managing the dynamism of the OSGi Service Platform in real-time Java applications.
\newblock Proceedings of the 27th Annual ACM Symposium on Applied Computing, ACM, 2012.
\newblock \doi{10.1145/2245276.2231952}

\bibitem{abt} F. Arbab.
\newblock Abstract Behavior Types: A Foundation Model for Components and Their Composition.
\newblock Formal Methods for Components and Objects.
\newblock vol. 2852 of LNCS, Springer-Verlag, 2003.
\newblock \doi{10.1007/978-3-540-39656-7\_2}

\bibitem{brada09} J. Bauml and P. Brada. Automated Versioning in OSGi: A Mechanism for Component Software Consistency Guarantee. 35th Euromicro Conference on Software Engineering and Advanced Applications, 2009.  
\newblock \doi{10.1109/SEAA.2009.80}

\bibitem{reportosgisem} J. O. Blech. Towards a Formalization of the OSGi Component Framework. http://arxiv.org/abs/1208.2563v1. arXiv.org 2012.


\bibitem{isolavision12} J. O. Blech, Y. Falcone, H. Rue{\ss}, Bernhard Sch{\"a}tz
\newblock Behavioral Specification based Runtime Monitors for OSGi Services.
\newblock Leveraging Applications of Formal Methods, Verification and Validation (ISoLA), 2012. 
\newblock \doi{10.1007/978-3-642-34026-0\_30}

\bibitem{blechschaetz12}
J. O. Blech, B. Sch\"{a}tz.
\newblock Towards a formal foundation of behavioral types for {UML}
  state-machines.
\newblock In: Proceedings of the 5th International Workshop {UML} and Formal
  Methods, 2012.
\newblock \doi{10.1145/2237796.2237814}


%\bibitem{abstractuml2005}
%F. S. de Boer, M. M. Bonsangue, M. Steffen und E. {\'A}brah{\'a}m. 
%\newblock A Fully Abstract Semantics for UML Components. 
%\newblock Formal Methods for Components and Objects, vol. 3657 of LNCS, Springer-Verlag,2005. 

\bibitem{webcontrmath} M. Bravetti, G. Zavattaro.
\newblock  A theory of contracts for strong service
compliance. Mathematical Structures in Computer Science 19(3): 601-638, 2009.
\newblock \doi{10.1017/S0960129509007658}

\bibitem{webcontracts} G. Castagna, N. Gesbert, L. Padovani. 
\newblock A theory of contracts for Web services. ACM
Trans. Program. Lang. Syst. 31(5), 2009.
\newblock \doi{10.1145/1538917.1538920}

\bibitem{plural} N. Cata{\~no} and I Ahmed. 
\newblock Lightweight Verification of a Multi-Task Threaded Server: A Case Study With The Plural Tool. Proceeding of Formal Methods for Industrial Critical Systems (FMICS), vol 6959 of LNCS, Springer, 2011.
\newblock \doi{10.1007/978-3-642-24431-5\_3}


\bibitem{jml}
P. Chalin, J.R. Kiniry, G.T. Leavens, E. Poll.
\newblock Beyond assertions: Advanced specification and verification with {JML} and {ESC/Java2}.
\newblock  Formal Methods for Components and Objects,  FMCO, vol. 4111 of LNCS, Springer 2005.
\newblock \doi{10.1007/11804192\_16}

\bibitem{cheng2011synthesis}
C. Cheng, H. Rue{\ss}, A. Knoll, C. Buckl.
\newblock Synthesis of fault-tolerant embedded systems using games: from theory
  to practice.
\newblock Verification, Model Checking, and Abstract Interpretation, vol.
  6538 of LNCS, Springer 2011.
\newblock \doi{10.1007/978-3-642-18275-4\_10}

\bibitem{ColacoPouzet:2003}
J.-L. Cola\c{c}o and M. Pouzet.
\newblock Clocks as first class abstract types.
\newblock  EMSOFT, vol. 2855 of LNCS, Springer, 2003.
\newblock \doi{10.1007/978-3-540-45212-6\_10}

\bibitem{conssercomp} J. L. Fiadeiro, A. Lopes. Consistency of Service
Composition. Fundamental Approaches to Software Engineering (FASE), vol. 7212 of LNCS, Springer, 2012.
\newblock \doi{10.1007/978-3-642-28872-2\_5}

\bibitem{osgisecuritybycontr}
O. Gadyatskaya, F. Massacci, A. Philippov.
\newblock Security-by-Contract for the OSGi Platform. 
\newblock Information Security and Privacy Conference, SEC, Springer, 2012.
\newblock \doi{10.1007/978-3-642-30436-1\_30}

\bibitem{brochjohnsen12} E. B. Johnsen and R. H\"ahnle and J. Sch\"afer and Rudolf Schlatte and Martin Steffen. ABS: A Core Language for Abstract Behavioral Specification. Post Conf. Proceedings 9th Intl. Symposium on Formal Methods for Components and Objects 2010. Springer-Verlag 2010.
\newblock \doi{10.1007/978-3-642-25271-6\_8}

\bibitem{muellerse2010} M. Mueller, M. Balz, M. Goedicke. Representing Formal Component Models in OSGi. Proc. of Software Engineering, Paderborn, Germany, 2010.

\bibitem{betypesptolemy}
E.A. Lee, Y.  Xiong.
\newblock A behavioral type system and its application in ptolemy ii.
\newblock Formal Aspects of Computing, 2004.
\newblock \doi{10.1007/s00165-004-0043-8}


\bibitem{osgi}
OSGi Alliance.
\newblock OSGi service platform core specification (2011) Version 4.3.

%\bibitem{papyrus}
%Papyrus UML.
%\newblock {\em {\url{http://www.papyrusuml.org}}}

\bibitem{behavioralrefinement} C. Prehofer. 
\newblock Behavioral refinement and compatibility of statechart extensions. Formal Engineering approaches to Software Components and Architectures. Electronic Notes in Theoretical Computer Science, 2012.

\bibitem{bip1}
J. Sifakis.
\newblock A framework for component-based construction -- {Extended Abstract}.
\newblock Software Engineering and Formal Methods, IEEE Computer Society, 2005.
\newblock \doi{10.1109/SEFM.2005.3}

\bibitem{mekontsotchinda:inria-00619233}
H.A.M. Tchinda,  N. Stouls, J. Ponge.
\newblock Sp{\'e}cification et substitution de services osgi.
\newblock Technical report, Inria (2011)
  \url{http://hal.inria.fr/inria-00619233}.


%\bibitem{uml}
%Unified Modeling Language (UML), Version 2.0.
%\newblock Object Management Group, August 2005.

\end{thebibliography}

\end{document}